\documentclass[prb,aps,
]{revtex4}
\usepackage{epsfig,epsf,psfrag}
\usepackage{graphicx}
\usepackage{epic,eepic}
\usepackage{color,pstcol}

\newcommand{\La}{\left\langle}
\newcommand{\Ra}{\right\rangle}
\newcommand{\Ls}{\left[}
\newcommand{\Rs}{\right]}
\newcommand{\Lf}{\left\{}
\newcommand{\Rf}{\right\}}
\newcommand{\Tr}{\rm Tr}

\begin{document}
\title{Non-equilibrium supercurrent through a quantum dot:
current harmonics\\ and proximity effect due to a normal metal lead}
\date{\today}

\author{T. Jonckheere$^{1}$}
\author{A. Zazunov$^{2}$}
\author{K.V. Bayandin$^{1,3}$}
\author{V. Shumeiko$^{1,4}$}
\author{T. Martin$^{1,5}$}

\affiliation{$^{1}$ Centre de Physique
Th\'eorique, Case 907 Luminy, 13288 Marseille cedex 9, France}
\affiliation{$^{2}$ Institut f\"ur Theoretische Physik, Heinrich-Heine-Universit\"at,
D-40225  D\"usseldorf, Germany}
\affiliation{$^{3}$ Landau Institute for Theoretical Physics RAS,
117940 Moscow, Russia}
\affiliation{$^{4}$ Department of Microelectronics and Nanoscience,
Chalmers University of Technology,
S-41296 G\"oteborg, Sweden}
\affiliation{$^{5}$ Universit\'e de la M\'edit\'erann\'ee, 13288
Marseille Cedex 9, France}

\date{\today}

\begin{abstract}
We study the out-of-equilibrium current through 
a quantum dot which is placed between two superconducting leads held at fixed voltage bias,
considering both cases of the absence and the presence of an additional normal lead 
connected to the dot.
Using the non-equilibrium Keldysh technique, 
we focus on the subgap bias region, where multiple Andreev reflections (MAR) are 
responsible for charge transfer through the dot. Attention is put on the DC current and on
the first harmonics of the supercurrent. 
Varying the position and/or the width of the dot level,
we first investigate the crossover between a quantum dot and
quantum point contact regimes in the absence of a normal lead.
We then study the effect of the normal electrode connected to the dot,    
which is understood to lead to dephasing, or alternatively to induce reverse proximity effect. 
By increasing the dot coupling to the normal probe,
we show the full crossover from zero dephasing to the incoherent case.
We also compute the Josephson current in the presence of the normal lead, and 
find it in excellent agreement with the 
values of the non-equlibrium current extrapolated at zero voltage.
\end{abstract}

\maketitle

\section{Introduction}

Non equilibrium transport between superconductors with a DC voltage bias
gives rise to a subgap structure in the current voltage
characteristics which can be described in terms of
Multiple Andreev Reflections (MAR)\cite{KBT,arnold}.
Indeed, it has been understood since the
sixties\cite{burstein} that when the bias potential between two
superconductors is smaller than the superconducting energy gap, electrons have to be
transferred in bunches in order to satisfy energy requirements.
The calculation of the current in the presence of
such MAR processes can proceed along several directions. Early
work\cite{KBT} considered a formulation of transport in terms of 
transmission probabilities rather than amplitudes.
During the last decade or so, MAR processes have been
studied theoretically in the coherent regime for point contacts,
using either scattering theory\cite{bratus,averin} or microscopic tight
binding Hamiltonians\cite{cuevas}. The coherent current which flows
between the two superconductors is then time dependent: it contains 
all harmonics of the Josephson frequency.  
Of particular interest in Ref. \onlinecite{averin,cuevas} was the fact that 
in addition to the DC current, the cosine and sine harmonics of 
the current were computed. These harmonics also exhibit structures
at the MAR onsets, and they allow for some additional diagnosis at 
low voltages: the amplitude of the sine first harmonic (SFH) at zero voltage
corresponds to the critical current in the Josephson (zero bias) limit, 
while the cosine first harmonic (CFH) vanishes. 

On the experimental side, in the context of mesoscopic physics,
pioneering experiments were performed on
atomic point contacts for the current\cite{mar_point_contacts_exp_current}
as well as for the noise\cite{mar_point_contacts_exp_noise}.
At the same time, samples containing a diffusive normal metal sandwiched between
superconducting leads were studied \cite{diffusive MAR_experiments}- a regime which
also triggered theoretical activity.~\cite{diffusive_MAR_theory}

Given the recent interest in nanophysics for studying systems with reduced size
(such as quantum dots and molecules), MAR through a
quantum dot described as a resonant level without interactions has
been addressed \cite{alfredo,johansson}. It was found that the position 
of the resonance bears strong consequences on the DC current voltage 
characteristics. So far however, little is known on the harmonics of the 
supercurrent. Experiments on non equilibrium supercurrent through
quantum dots have recently been performed, pointing out the
effect of size quantization and in some cases of resonances, attributed to
Coulomb interactions\cite{schonenberger,experiments_MAR_ quantum_dots}. 

So far MAR transport calculations have been focused either on the
incoherent or coherent cases. The crossover regime was discussed 
for a chaotic dot junction where an external magnetic field 
serves as a dephasing factor\cite{Samuelsson2002}. In mesoscopic physics, the
current (and noise) which flows between a source and a drain can
be modified if one inserts a voltage probe between the
two\cite{buttiker_dephasing,beenakker_buttiker}. The probe voltage
can be adjusted so that no net electrons flow through it,
electrons which enter this probe loose their phase coherence
because they enter in contact with an electron reservoir. In
mesoscopic transport, such probes can be used as a way to mimic
the effect of inelastic scattering/decoherence on transport
between the source and the drain. One can thus model the crossover
between the two regimes by adjusting
the degree of decoherence by varying the coupling to such voltage
probes. At the same time, experimentally one could also construct
actual nanostructures which contain a controlled connection to
such voltage-like probes.

In this paper, we address key issues associated with MAR for a device 
consisting of two superconductor connected to a quantum dot/resonant level,
in the absence and in the presence of a normal metal lead connected to the dot. 
To achieve this, we will choose a Hamiltonian formulation and 
we will use a Keldysh Green's function approach where the leads are
effectively integrated out.
First, we will study the supercurrent harmonics of an isolated quantum dot. 
We will first show under which conditions the regimes of a point contact and of a resonant level 
are recovered, and then discuss the properties of the supercurrent harmonics for the case 
of the resonant level; we will
show that these properties differ strongly from those of a quantum point contact. 
Next, we will introduce a normal probe which is directly 
connected to the dot, and we will monitor the full crossover from the coherent 
regime to the fully incoherent regime. We will also compute the 
(equilibrium) Josephson current independently, and we will check whether its first harmonic
 corresponds to the extrapolated amplitude of the SFH at zero bias. 
Note that the setup is similar to that of Ref. \onlinecite{konig},
however, the attention there was put at the non-equilibrium Josephson effect 
under current injection from the normal probe.

Note that the inclusion of a normal probe can be interpreted as a source 
of dephasing\cite{buttiker_dephasing,beenakker_buttiker},
extended to the case of a supercurrent. Yet it can also have an alternative
interpretation. In the presence of superconducting leads, superconducting 
correlations are induced on the dot because of the (usual) proximity effect. 
This, for instance is illustrated in several recent works both in equilibrium\cite{zazunov_feinberg}
and in voltage biased\cite{zazunov_egger} superconducting devices where 
an effective action for the dot is derived.    
Now, the addition of a normal lead will induce a reverse proximity effect 
on the dot which will compete with the existing one: superconducting correlations
on the dot will be suppressed.  It is the ratio between 
the escape rates to the superconducting leads and to the normal leads 
which will determine the coherent/incoherent character of transport. 

The paper is organized as follows. In Sec. II, the general model is introduced.
Sec. III reviews the non-equilibrium formalism to study transport.
Details on how the current harmonics are computed are presented in Sec. IV.
The DC current and harmonics are discussed in Sec. V for several limiting cases
and in the presence of the normal lead,
and the comparison with the Josephson current calculation is performed in Sec. 
VI. We conclude in Sec. VII, while details about the numerical implementation are
given in Appendix A.

\section{model Hamiltonian}
\label{model_hamiltonian}
\begin{figure}
\centerline{\includegraphics[width=8cm]{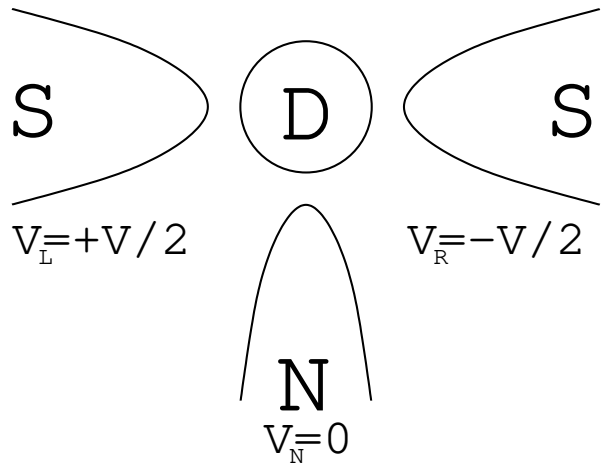}}
\caption{Schematic picture of the setup. A single-level quantum
dot is placed between two superconductors, which are biased with
voltages $\pm V/2$. An additional normal metal lead plays the role
of a voltage probe\cite{buttiker_dephasing}. Its coupling to the
dot is controllable and its voltage is taken to satisfy a
zero-current condition for the probe.\label{fig00}}
\end{figure}

The system we study consists of a quantum dot attached
to two superconducting leads. For simplicity, the coupling
between the two leads is assumed to be symmetric.
In addition, a third lead in the normal metal regime is attached
to the dot (Fig. \ref{fig00}). This model is a generalization
of the two terminal structure presented in Refs.
\onlinecite{zazunov_egger}. 
The tunnel coupling from
the dot to the normal lead will be a control parameter
for adjusting dephasing effects. The dot contains for simplicity a
relevant single electronic level, but the analysis can be
straightforwardly extended to treat a multilevel dot. We label the
applied bias voltage on the left (right) lead $V_L$ ($V_R$), where
$V_j$ ($j=L,R$) is measured with respect to the dot level (see the
tunnel Hamiltonian below). The chemical potential of the normal
lead is set at $V_N=0$.

We focus on the most interesting
low temperature ($T$) limit with moderate (but not low)
transmission to the superconducting electrodes, which allows in principle to
neglect Coulomb charging effects.
The resulting Hamiltonian of the system with the time-dependent
tunnelling terms reads:
\begin{equation}
H = H_D + \sum_{j=L,R,N} H_j + H_T(t) ~,
\end{equation}
where
\begin{equation}
H_{D} = \epsilon_0 \, \sum_{\sigma = \uparrow, \downarrow} d^\dagger_\sigma d_\sigma ~.
\label{dot_hamiltonian}
\end{equation}
%
The BCS Hamiltionian of the superconductors and normal metal is expressed in terms of Nambu spinors:
\begin{equation}
H_j = \sum_k \Psi^\dagger_{jk} \left(
\xi_k \, \sigma_z + \Delta_j \, \sigma_x \right) \Psi_{jk} ~,~~~
\Psi_{jk} = \left(
\begin{array}{c}
\psi_{jk, \uparrow} \\
\psi^\dagger_{j(-k), \downarrow}
\end{array} \right) ~,~~
\xi_k = k^2 / (2 m) - \mu ~,
\end{equation}
with $\sigma_z$, $\sigma_x$ Pauli matrices in Nambu space. The gap
is assumed to be the same for the two superconducting leads
($j=L,R$) $\Delta_j\equiv \Delta$, while for the normal lead
($j=N$) $\Delta_N\equiv 0$. With the Nambu notation, the
tunnelling term becomes:
\begin{equation}
H_T(t) = \sum_{jk} \,
\Psi^\dagger_{jk} \, {\cal T}_j(t) \, d + {\rm h.c.} ~,~~~
d = \left(
\begin{array}{c}
d_\uparrow \\ d^\dagger_\downarrow
\end{array} \right) ~,
\label{H_T}
\end{equation}
with a tunneling amplitude ${\cal T}_j(t) = t_j \sigma_z \, e^{i
\sigma_z \chi_j(t)/2}$, where the presence of the bias voltage induces a
time dependence of the extracted
phases of the superconducting order parameters:
$\chi_j(t) = \sigma_j \,t V_j/2$, $\sigma_j = \pm 1$  for $j = L/R$.We often set $\hbar = e=1$ and restore them when convenient in final expressions.
With these notations, the current operator between lead $j$ and the dot reads:
\begin{equation}
I_{j}(t)=i \sum\limits_{k}\Psi_{jk}^{\dagger}\sigma_{z}{\cal
T}_{j}(t)d(t)+h.c.
\end{equation}


\section{Keldysh formalism}

We use the Keldysh  Green's function formalism in order to compute
the current. The details of such calculations have appeared elsewhere\cite{zazunov_egger},
so here we only summarize the techniques. The Coulomb interaction is neglected here.
First, a Keldysh partition function is introduced [$H_0 = H_D +\sum_{j=L,R,N} H_j$]:
\begin{equation}
Z=\Tr\Lf{}e^{-\beta{}H_{0}}S(\infty,\eta)\Rf ~,
\end{equation}
where
\begin{equation}
S(\infty,\eta)=T_{c}\exp\Lf-i\int_{-\infty}^{+\infty}\!\!dt\,{\cal
H}_{T}(t,\eta)\Rf,~
\end{equation}
and
\begin{equation}
{\cal
H}_{T}(t,\eta)=\sum\limits_{jk}\hat\Psi_{jk}^{\dagger}{\cal
 T}_{j}(t)\hat\tau_{z}e^{i\hat\tau_{z}\sigma_{z}\eta_{j}(t)/2}\hat{d}(t)+h.c.,
\end{equation}
$\tau_{z}$ is a Pauli Matrix in Keldysh space, and we have introduced a
counting field $\eta_j$ for each lead-dot coupling,
which allows to compute the current by deriving the partition
with respect to it:
\begin{eqnarray}
\La I_{j}(t)\Ra=iZ_{0}^{-1}\left.\frac{\delta
Z[\eta(t)]}{\delta\eta_{j}(t)}\right|_{\eta=0} ~.
\end{eqnarray}
The action possesses a quadratic dependence on the lead
fermion spinors, so the latter can be integrated out\cite{zazunov_egger}. One is left with
an action which depends on the dot spinors only, for which the effect
of the leads appears via self energies in Keldysh Nambu space:
\begin{eqnarray}
\hat{\Sigma}_{j}(t_{1},t_{2})=\Gamma_{j}\int_{-\infty}^{\infty}\frac{d\omega}{2\pi}
~e^{-i\omega(t_{1}-t_{2})}
e^{-i\sigma_{z}V_{j}t_{1}}[\omega\cdot{\bf 1}-\Delta_{j}\cdot\sigma_{x}]~
e^{i\sigma_{z}V_{j}t_{2}} \nonumber \\
\times \,
\left[-\frac{\Theta(\Delta_{j}-|\omega|)}{\sqrt{\Delta_{j}^{2}-\omega^{2}}}\hat{\tau}_{z}+i\,{\rm
sign}(\omega)\frac{\Theta(|\omega|-\Delta_{j})}{\sqrt{\omega^{2}-\Delta_{j}^{2}}}\left(\begin{array}{cc}
  2f_{\omega}-1 &~~ -2f_{\omega} \\
  +2f_{-\omega} &~~ 2f_{\omega}-1 \\
\end{array}\right)\right],\label{eq05}
\end{eqnarray}
where we define the escape rates from the dot as
\begin{equation}
\Gamma_{j}=\pi \nu_{j}(0) \left|t_{j}\right|^{2} ,
\label{eq:Gamma}
\end{equation}
 with $\nu_j(0)$ the (constant) density
of states of lead $j$ in the normal state. Fermi filling factors
are given by $f_{\omega}=1/(e^{\beta\omega}+1)$. 
With this definition, the formula for the partial average current reduces
to\cite{zazunov_egger}:
\begin{eqnarray}
\La
I_{j}(t)\Ra &=& \frac{1}{2}~\Tr\Lf\hat\tau_{z}\sigma_{z}\int_{-\infty}^{+\infty}dt^{\prime}
\left(\hat{G}(t,t^{\prime})\hat{\Sigma}_{j}(t^{\prime},t)-\hat{\Sigma}_{j}(t,t^{\prime})\hat{G}(t^{\prime},t)\right)\Rf \nonumber \\&=&
-2{\rm{}Re~}{\rm
tr}\Lf\sigma_{z}\left(\hat{\Sigma}_{j}\circ\hat{G}\right)^{\!\!+-}\!\!\!\!\!\!(t,t)\Rf ~,
\label{mw}\end{eqnarray} 
where the trace ``${\rm tr}$'' operates
only in Nambu space and the symbol $\circ$ denotes convolution in
time. Here the Green's function of the dot (which is dressed by
the leads) is defined as:
\begin{equation}
G_{\alpha\beta}^{ss^{\prime}}(t,t^{\prime})=-i\La T_{c}\Lf
S(\infty)d_{\alpha}^{s}(t)d_{\beta}^{\dagger
s^{\prime}}(t^{\prime})\Rf\Ra_{0} ~.
\end{equation}
This constitutes an extension of the Meir-Wingreen\cite{mw} formula for a dot connected to one or several superconducting
or normal metal leads. Provided that the Green's function is computed exactly it applies also for situations where there are interactions (electron-electron or electron-phonon) on the dot.

\section{Calculation of the MAR current}

We are interested in the calculation of the electrical current
through a quantum dot with a single level $\epsilon_0$, which is
placed near two superconducting leads with applied voltages
$V_{L}=V/2$ and $V_{R}=-V/2$. The reference position for the zero
of energy is chosen halfway between the two chemical potentials of
the superconductors. In principle, the position of the dot level
should depend on the geometry (escape rates) and on the applied
bias, and it could be derived in a self-consistent way. Here, 
we ignore such dependence and assume that the dot level
position can be changed using a metallic gate located close to the dot.
Nevertheless, an important part of our study will deal with 
$\epsilon_0\equiv 0$. In order to avoid the proliferation of
parameters, we choose to specify a symmetric device, where the
transparencies of the left and right superconducting leads are the
same, while no assumption is made about the normal lead.
With this choice, at $\epsilon_0=0$ the requirement that no net current flows
in the normal leads imposes that $V_{N}=0$ invoking electron/hole symmetry. 
In what follows, all energy/frequency scales are expressed in units of the
superconducting gap $\Delta$.

When a bias is applied to superconductor, the current is not stationary
as it contains all harmonics at the Josephson frequency $\omega_J=2eV/\hbar$.
It is then convenient to introduce a double Fourier transform with
summations over discrete domains in frequency:
\begin{eqnarray}
G(t,t^{\prime})&=&\sum\limits_{n,m=-\infty}^{+\infty}\int_{F}\frac{d\omega}{2\pi}e^{-i\omega_{n}t+i\omega_{m}t^{\prime}}G_{nm}(\omega)~,
\label{eq:doublefourier1}\\
\Sigma(t,t^{\prime})&=&\sum\limits_{n,m=-\infty}^{+\infty}\int_{F}\frac{d\omega}{2\pi}e^{-i\omega_{n}t+i\omega_{m}t^{\prime}}\Sigma_{nm}(\omega)~,
\label{eq:doublefourier2}
\end{eqnarray}
where $\omega_{n}=\omega+nV$, and the frequency integration is performed over a
finite domain $F\equiv [-V/2,V/2]$, where $V$ is the
voltage on the leads. With these definitions, the Fourier transform of the current becomes:
\begin{equation}\!\!\!\!\!\!\!\!
I_{j}(\omega^{\prime})=\!\!\sum\limits_{n,l}\!\!~2\pi\delta\Big(\omega^{\prime}-(n-l)V\Big)
\frac{1}{2}\int_{F}\frac{d\omega}{2\pi}\Tr\Lf\sigma_{z}\hat\tau_{z}
\!\!\sum\limits_{m}\!\!\left[\hat{G}_{nm}(\omega)\hat\Sigma_{j,ml}(\omega)-
\hat\Sigma_{j,nm}(\omega)\hat{G}_{ml}(\omega)\right]\Rf,\label{eq01}
\end{equation}
where the matrices $\hat{G}_{nm}(\omega)$ and
$\hat\Sigma_{j,nm}(\omega)$ have a $4\times{}4$
block structure for every pair of energy domain indexes $n$ and $m$.

In the double Fourier representation, the lead self-energy is given by\cite{zazunov_egger}:
\begin{equation}
\hat{\Sigma}_{j,nm}(\omega_{n})=\Gamma_{j}\left[\begin{array}{cc}
  \delta_{n,m}\hat{X}_{j}(\omega_{n}-V_{j}) & \delta_{n-2V_{j}/V,m}\hat{Y}_{j}(\omega_{n}-V_{j}) \\
  \delta_{n+2V_{j}/V,m}\hat{Y}_{j}(\omega_{n}+V_{j}) & \delta_{n,m}\hat{X}_{j}(\omega_{n}+V_{j}) \\
\end{array}\right]~,
\end{equation}
where
\begin{equation}
\hat{X}_{j}(\omega)=\left[-\frac{\Theta(\Delta_{j}-|\omega|)\omega}{\sqrt{\Delta_{j}^{2}-\omega^{2}}}\hat{\tau}_{z}
+i\frac{\Theta(|\omega|-\Delta_{j})|\omega|}{\sqrt{\omega^{2}-\Delta_{j}^{2}}}\left(\begin{array}{cc}
  2f_{\omega}-1 &~~ -2f_{\omega} \\
  +2f_{-\omega} &~~ 2f_{\omega}-1
\end{array}\right)\right],
\end{equation}
and $\hat{Y}_{j}(\omega)=-\Delta_{j}\hat{X}_{j}(\omega)/\omega$.

The dressed Green's function for the dot appears from the Dyson
equation of the form:
\begin{equation}
\hat{G}_{nm}(\omega)=\Ls\hat{G}_{0,nm}^{-1}(\omega)-\hat{\Sigma}_{T,nm}(\omega)\Rs^{-1}~,\label{inversion_dyson}
\label{eq:dressedG}
\end{equation}
where
$\hat{G}_{0,nm}^{-1}(\omega)=\delta_{nm}(\omega_{n}-\epsilon_0\sigma_{z}) \hat\tau_{z}$
and $\hat\Sigma_{T}=\sum\limits_{j}\hat\Sigma_{j}$.

The dressed Green's function is obtained numerically: in practice, it requires
the inversion of a ``large'' matrix (Eq.~(\ref{inversion_dyson})). This finite matrix
is obtained by limiting the discrete Fourier transforms (Eqs.~(\ref{eq:doublefourier1}) and (\ref{eq:doublefourier2})) to
a cut-off energy $E_c$, which has to be large compared to all the relevant energies of the problem.
 This energy $E_c$ defines a finite number of frequency domains $n_{max}$. 
As the width of each domain is $\sim V$, one has $n_{max} \sim 1/V$, which reflects the fact that at
small voltages, one needs to sum on a very large number of Andreev reflections. In practice, we have
chosen in most case $E_c=16 \Delta $, and we have computed the current for $V>0.1 $ only. See appendix~\ref{sec:appendix}
for more details on the numerical implementation.

\section{Results for the MAR current}
\label{results_for non_interacting dot_section}

We present here the results for the MAR currents in different regimes.
First, in sections~\ref{subsec:A} and \ref{resonant_section}, we consider the case without coupling to a normal lead,
respectively in the quantum point contact (QPC) limit and the the resonant level regime. This will allow us to compare
with existing results, and also to provide new results for the resonant level case.
Then, in  section~\ref{Mar_gammaN_section}, we consider the effect of the coupling to the normal lead on the 
MAR current.
The current is plotted in units of $e\Delta/h$ and  calculations are performed
choosing a temperature $\theta=0.01\Delta$ in order to study the
bias dominated regime. 

\subsection{Quantum point contact limit}
\label{subsec:A}

\begin{figure*}
\centerline{\includegraphics[width=17cm]{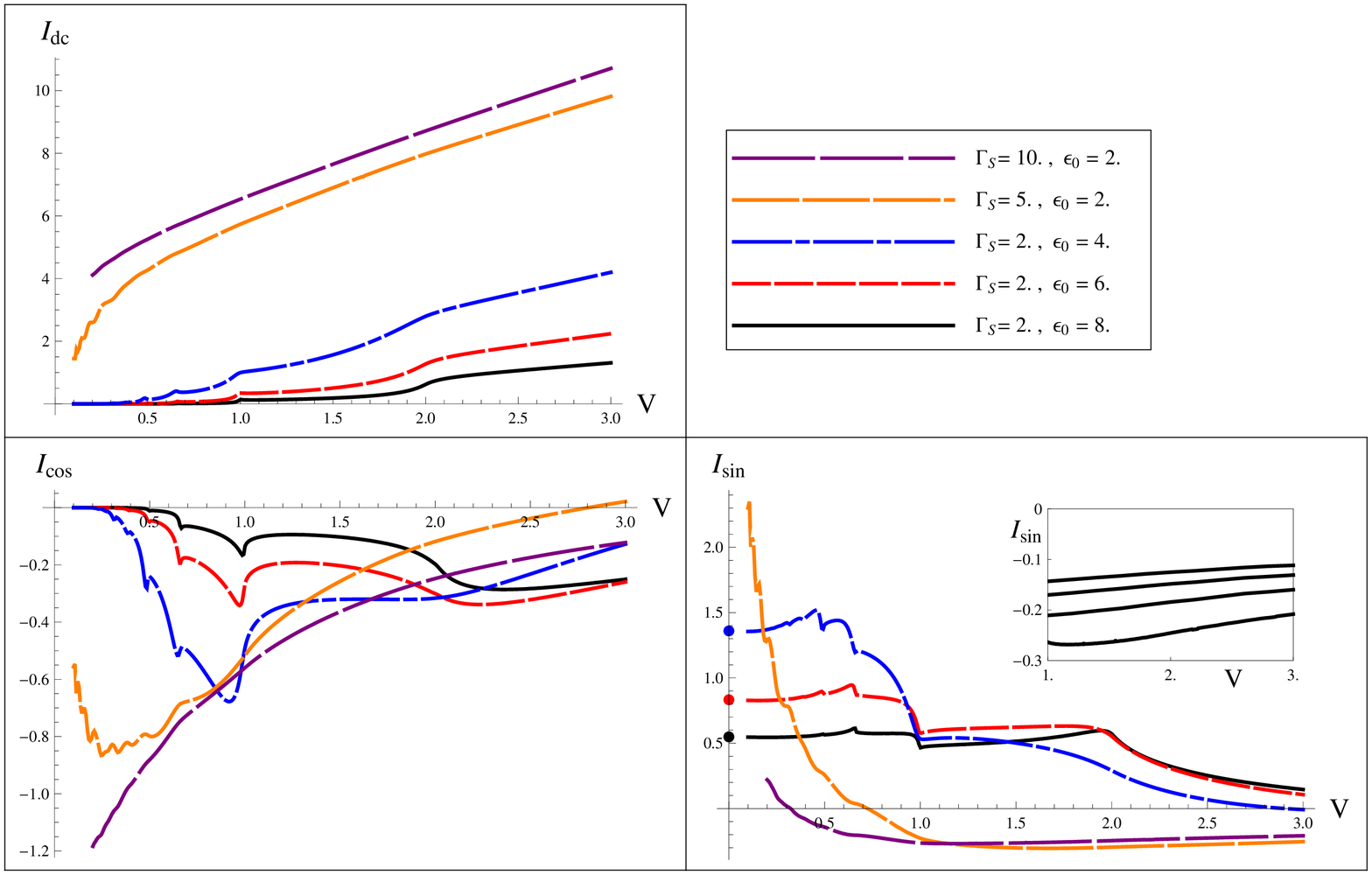}}
\caption{(Color online) Quantum Point Contact limit.
DC (top panel), cosine first harmonic (lower left panel), and 
sine first harmonic (lower right panel) of the MAR current 
for different escape rates and dot level position. The circles on the $y$
axis for the sine harmonic show the value of the first harmonics of the 
equilibrium Josephson current (computed independently). The inset in the sine
harmonics panel shows results for $\Gamma_S=10, 20, 30, 40$ (bottom to top)
for $V$ between 1.0 and 3.0.
\label{figure2}}
\end{figure*}

If the escape rate $\Gamma_{S} \equiv \Gamma_{L}=\Gamma_{R}$, see Eq.~(\ref{eq:Gamma}), is sufficiently large compared to the
superconducting gap, the resulting open quantum dot sandwiched between
the superconducting electrodes should behave like a point contact.
The case of a point contact was studied both in the context of 
scattering theory\cite{bratus,averin} and from a Hamiltonian approach\cite{cuevas}
a decade ago. 
There, results were obtained both for the DC current as well as for the first harmonics 
 $I_{cos}$ and $I_{sin}$ of the current.
\begin{equation}
I(t)=I_{dc}+I_{cos}\cos(2eVt/\hbar)+I_{sin}\sin(2eVt/\hbar)+\dots
\end{equation}
Fig. \ref{figure2} displays these quantities for the case of a open quantum dot, 
for several escape rates and for several level positions of the dot. 
The challenge lies with the fact that in order to reproduce the case of the QPC at high transmission (large coupling $\Gamma_S$), 
one has to include a large number of Andreev reflections between the dot and the superconducting leads, 
especially at low voltages. In practice, this means using a large cut-off energy $E_c$ and thus a very large matrix size $\sim n_{max}$
for small voltage. The existing results for the DC current as well as for the 
first harmonics\cite{cuevas,averin} provide a reference for our results in the QPC limit. 

In the top panel of Fig.~\ref{figure2}, we plot the
DC current for several values of $\Gamma_{S}>\Delta$, and several 
values of the dot level position, which allow to describe the crossover from the 
high transmission to the low transmission regime. 
In the normal state, the transmission
probability between the source and drain electrodes as a function of energy $\epsilon$
is given by a Lorentzian line shape: 
\begin{equation}
T(\epsilon)=\Big[ 1+(\epsilon-\epsilon_0)^2/\Gamma_S^2 \Big]^{-1} ,
\label{lorentz}\end{equation}
and this transmission is unity (resonant behavior) whenever $\epsilon-\epsilon_0=0$.
This resonant behavior in principle affects the transmission of electrons and holes
through the quantum dot.  
In our calculations to describe the QPC limit, $\epsilon_0$ is shifted 
away from zero beyond the superconducting gap, because otherwise, regardless of the 
line width broadening provided by $\Gamma_S$, there are reminiscences of
resonant MAR behavior (for a description of these resonant processes, see Sec. \ref{resonant_section}). 
For $\epsilon_0=2$ and $\Gamma_S=5\Delta$ or $\Gamma_S=10\Delta$, no significant features of MAR are found in the DC current, in accordance 
with Refs. \onlinecite{averin,cuevas}. The DC current is linear for voltages larger 
than the quasiparticle onset $V>2\Delta$, and it vanishes abruptly close to $V\sim 0.1\Delta$, 
where some MAR structure is noticeable for $\Gamma_S=5$ (lower transparency). Note that we have
used increased cut-off energies ($E_c=30$ and $50$) for the cases $\Gamma_S= 5$ and $10$ respectively; this limits 
the data for $\Gamma_S=10$ to $V>0.2$ . In order to further reduce the transparency 
and thus observe some structure, one can either decrease the escape rate or 
drive $\epsilon_0$ further away from $0$ . Cusps in the current derivative 
appear as expected at the so-called MAR onsets defined as $eV=2\Delta/n$, $n$ integer,
($\Gamma_S=2$, $\epsilon_0=4$).
As the transparency is reduced ($\Gamma_S=2$, $\epsilon_0=6$ and  $8$) only the quasiparticle 
onset ($n=1$) and the $n=2$ onset are identifiable. By reproducing the calculations 
of Ref. \onlinecite{averin}, we have compared quantitatively our results with
those for a true QPC, and found that, by using the transparency $T$ of the QPC as a fitting
parameter, we can get a near perfect agreement between the two results (not shown). For 
example, the values of $T$ corresponding to $\Gamma_S=2$, $\epsilon = 4,6$ and $8$ are
respectively $T=0.55,0.35$ and $0.23$.  

In the bottom panels of Fig. \ref{figure2}, the first harmonics are displayed. 
We first discuss their general behavior. 
Both the cosine and sine harmonics display structures at the MAR onsets, in the same 
manner as the DC current. These structures are more pronounced for lower transparencies 
(but note that for $\Gamma_S=5\Delta$, $\epsilon_0=2$ some structure
is found for onsets corresponding to $n>3$).
For voltages beyond the single particle current ($n=1$) onset, the 
cosine and sine harmonics eventually decay.
The cosine harmonic vanishes at low voltage, which is clear for all parameters 
displayed, except for the one which corresponds to the highest transmission ($\Gamma_S=10$)
because of our numerical limitation for $V\to 0$ (i.e. the vanishing of the cosine harmonics
for $\Gamma_S=10$ should be visible for $V<0.2$).  
The sine harmonics saturates at a non zero, positive value, for $V \to 0$. This is a signature of the fact that the Josephson effect
operates at zero voltage, and our non equilibrium calculation at $V\to 0$ should in principle match with 
a calculation of the (equilibrium) Josephson current. The value of the first harmonics of the Josephson current,
computed independently (see section~\ref{josephson_section}), is shown by a circle on the $y$ axis for the low transmission curves.
One can see that there is an excellent agreement between this value and the extrapolation of the non-equilibrium data for
$V \to 0$ (quantitatively, the agreement is better than 1 \%).   
As in the case of the DC current, we have compared these results with 
those obtained for a QPC in Ref. \onlinecite{averin}. 
Using the values of the QPC transmission obtained with the DC current fits,
we find a good qualitative agreement between the two results for the cos
component, except at large voltage where the results we obtained decrease (in absolute value) much faster.  For the 
sin component, the shapes of the curves are similar, but there is an overall downwards shift of the our results 
with respect to the QPC results. We attribute this to the fact that although the dot level has been positioned away from the 
superconducting gap region, the energy dependent transmission of the dot still 
plays a role in the MAR processes, and a perfect match with the QPC limit is not attainable.
Note that we also find a downwards shift of the first harmonics of the Josephson current with respect to the first
harmonics of the Josephson current of a real QPC\cite{averin}, using the transmission values found with the DC fits.  
Finally, the fact that the sine harmonic is negative for large $V$ ($V>\Delta$) 
at very large $\Gamma_S$ ($\Gamma_S=10$ or $5$, $\epsilon_0=2$)
remains a puzzle for us, but it seems to be related to this global
downards shift.
 We have carefuly checked that this is 
not due to any numerical imprecision or convergence problem (see Appendix A). 
The inset on the sine harmonics plot shows the results at large $V$ 
for larger values of $\Gamma_S$ ($\Gamma_S=10, 20, 30, 40$ from bottom to top). It shows that even with larger $\Gamma_S$ the sine harmonics
remains negative at large $V$, however it tends to decrease in absolute value as $\Gamma_S$ is increased.

\subsection{Narrow resonance limit}   
\label{resonant_section}

We now shift the discussion to the case where the line width of the dot 
is smaller than the superconducting gap, setting for convenience 
the dot level position at $\epsilon_0=0$. The DC current was studied
previously for similar parameters\cite{alfredo,johansson}. Here
one of our aims is to also include a discussion of the harmonics 
of the current which is absent in the literature. 

\begin{figure*}
\centerline{\includegraphics[width=17cm]{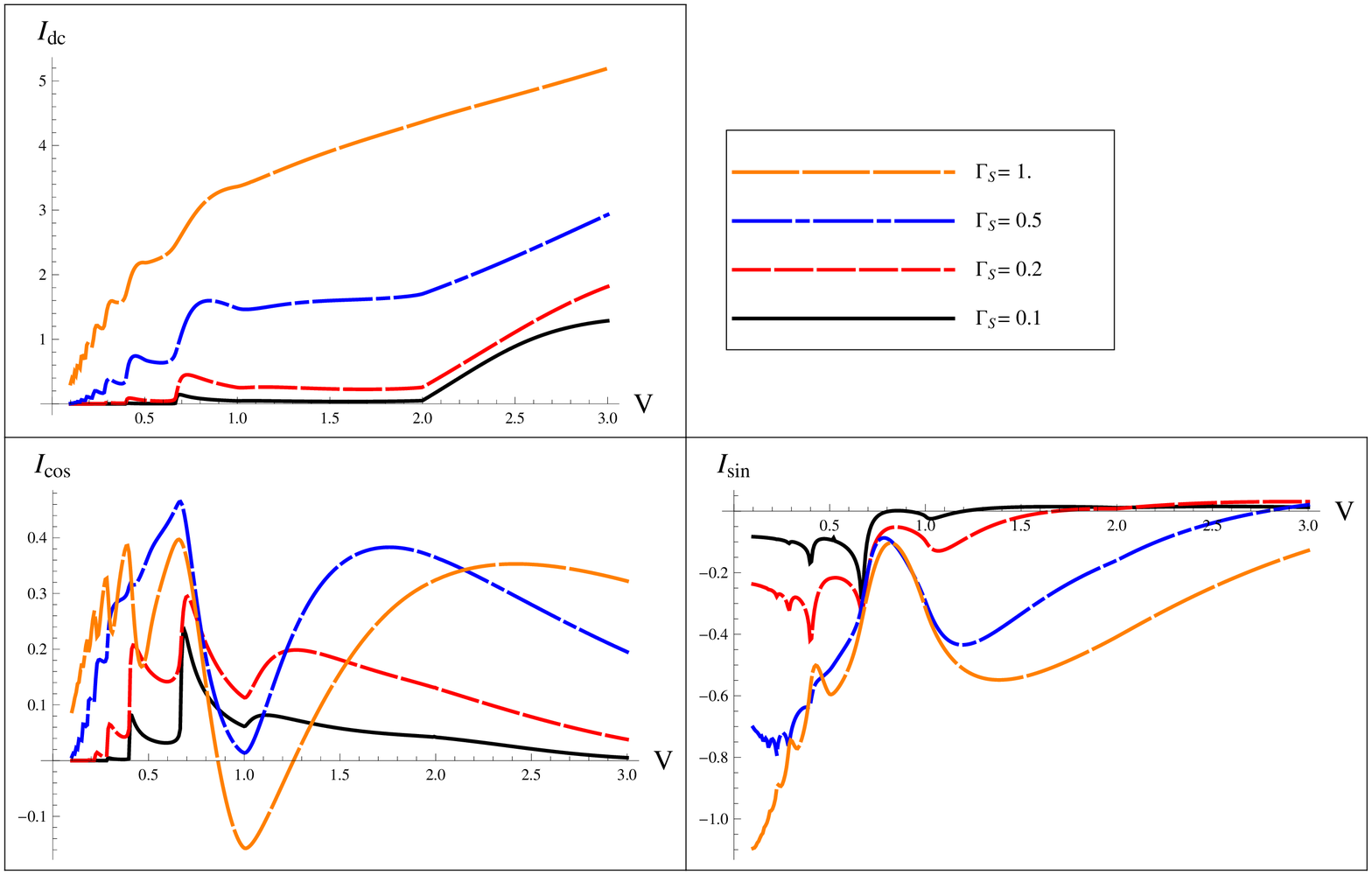}}
\caption{
(Color online) Resonant level regime. DC (top panel), cosine first harmonic (lower left panel), and 
sine first harmonic (lower right panel) of the MAR current, with dot level position $\epsilon_0=0$ and  different escape rates $\Gamma_S$. 
\label{figure3}}
\end{figure*}

In the top panel of Fig.~\ref{figure3} we show that we are able
to reproduce these results for the DC-current.  One
finds a very good agreement with existing results: 
tunnel Hamiltonian approach to all orders, or , alternatively 
scattering theory approach with the Lorentzian line shape of Eq. (\ref{lorentz}). 
For the DC current, the striking effect of the presence of 
a resonance is the fact that it favors specific Andreev 
reflection processes. Indeed, some structure is found at 
the odd $n$ onsets (recall that the onsets are located at $eV=2\Delta/n$). In the 
so called MAR ladder picture, an electron which transits from 
one superconductor to the other gains an energy $eV$, and the 
same is valid for a hole which is reflected back. 
Because of the presence of the resonance the 
trajectories of electrons and holes will have larger or smaller weights in
the DC current: a resonant trajectory
corresponds to the situation where the electron or hole 
energy crosses at one point the resonant level\cite{alfredo,johansson}.
For the case of $n$ odd, with the resonant level located 
in the middle of the bias window, a resonant
hole trajectory always occurs after $(n-1)/2$ reflections. 
For instance, at $n=3$
the first and last electron trajectory are non resonant, while 
the hole trajectory is. 
We have also observed (not shown here) that shifting the position 
of the resonant level  leads to an overall shift of the 
MAR structure. Displacing the resonance means that the electron trajectories
around the $(n\pm 1)/2$ reflection will have an enhanced transmission
while the hole resonance weakens. 

In the top panel of Fig.~\ref{figure3}, the DC current for several values of
 the resonance line width are presented. As expected, nothing significant occurs for the DC current
with decreasing $\Gamma_S$: the MAR structure gets sharper, but on the 
other hand the DC current is reduced. For voltages beyond the superconducting 
gap, and substantial line widths $\Gamma_S=0.5, 1.$, the behavior 
seems to be linear, while for $\Gamma_S=0.1, 0.2$ we see the beginning of a saturation 
for the current at large voltages. Indeed, in the limit $eV\gg \Delta$
the current is only due to quasiparticle transfer and it 
behaves like the current of a resonant level between two normal leads.
The apparent lack of saturation for  $\Gamma_S=0.5, 1.$ only reflects 
the fact that we are displaying a limited voltage range in order to focus on the 
MAR structure.       

We turn now to the first harmonics of the current.
Both harmonics decay at voltages beyond the superconducting gap
(quasiparticle dominated regime, as explained above). 
MAR onsets appear at the odd onsets
found in the DC current. However, the cosine 
harmonic (Fig. \ref{figure3} lower left panel)
shows a dip at the $n=2$ onset. It vanishes 
at low voltage, and displays large voltage oscillations
in this range. While the cosine harmonic was strictly negative for 
the QPC limit, here we see that its sign is reversed, 
except for the fact that it takes some negative values 
at relatively high transparencies ($\Gamma=1.$) near $eV=\Delta$. 

The sine harmonic (Fig. \ref{figure3} lower right panel)
also has a reversed sign with respect to the QPC situation for the voltage 
range under study, except perhaps at high voltages ($eV>2\Delta$) and high 
transparencies. As the voltage is lowered, the sine harmonic decreases
in amplitude and seems to saturate at some negative value when the voltage approaches zero. 
One would expect that the SFH would approach the stationary Josephson current value when $V=0$, as it is in a point contact
\cite{averin,bratus1997}. However, this would mean in the present case, that the junction is in the $\pi$-junction regime, 
which is only possible either under strong Coulomb interaction\cite{glazman_matveev,spivak_kivelson}
or in spin active junctions\cite{benjamin,michelsen2008}.
In non-interacting non-magnetic resonant contacts the Josephson current 
is always positive\cite{beenakker_van_houten,wendin1996}.  
This phenomenon can be explained with strong non-equilibrium population of the Andreev levels in the MAR regime,
which changes their contribution to the current compared to the equilibrium regime.
 At small bias voltage, the MAR can be understood in terms of adiabatically moving Andreev levels
 and Landau-Zener transitions between the levels, and between the levels and continuum states\cite{averin,bratus1997}.
 In point contacts, the Andreev level positions are strongly asymmetric with respect to the chemical potential 
(except at the full transmission limit): one level lies below the chemical potential 
close to the filled continuum state band, while another lies above the chemical potential close to the depleted band. 
At small voltage the quasiparticles exchange between the levels is negligible, 
and the level population is determined by an exchange with the corresponding continuum band, 
thus keeping level populations close to equilibrium. In resonant contacts the situation is different: 
when $\epsilon_0=0$, the levels are symmetric with respect to the chemical potential,
 they interact with the continuum bands identically, and therefore their populations are identical
(neglecting weak effect of inelastic relaxation). This eventually switches off the level contribution 
to the Josephon current, revealing current of the continuum states, which is negative\cite{beenakker_van_houten}.
In equilibrium, the Josephson current is dominated by the Andreev levels, and it is positive. 
The effect should gradually vanish when the resonant level shifts from the chemical potential, 
because the Andreev level positions become asymmetric as in the point contacts. 
We note that similar discontinuity of SFH at zero voltage exists also in fully transparent point contact.  

The presence of the resonant level thus modifies drastically 
the MAR current. It changes the DC current voltage characteristics 
by favoring some structure at odd MAR onsets. 
For the sine and cosine harmonics, the effect is more dramatic 
as these harmonics change sign in (almost) the whole range of the 
subgap voltage. 

\subsection{MAR current in the presence of a normal lead}
\label{Mar_gammaN_section}

We now turn to the central point of this study: how does the 
gradual coupling to a normal lead affect the MAR current? 
First, the dot level is set at $\epsilon_0=0$, and the 
chemical potential of the normal lead $\mu_N=0$, in order 
to insure that no net current flows through this lead. 
We focus in this section only on the case of the resonant level
regime $\Gamma_S=0.2$ rather than the point contact regime
$\Gamma_S>2$. A reason for this choice is first the fact
that we expect the dot level to be broadened by the 
presence of the additional lead, and in order to 
observe any broadening effect we need to start with a 
rather sharp level. Second, we have seen in Sec. 
\ref{resonant_section} that the presence of a narrow 
resonance gives rise to rather sharp features in the 
DC current and its harmonics, so it is natural to ask 
how these features are modified by the presence of the
normal lead.   

\begin{figure*}
\centerline{\includegraphics[width=16cm]{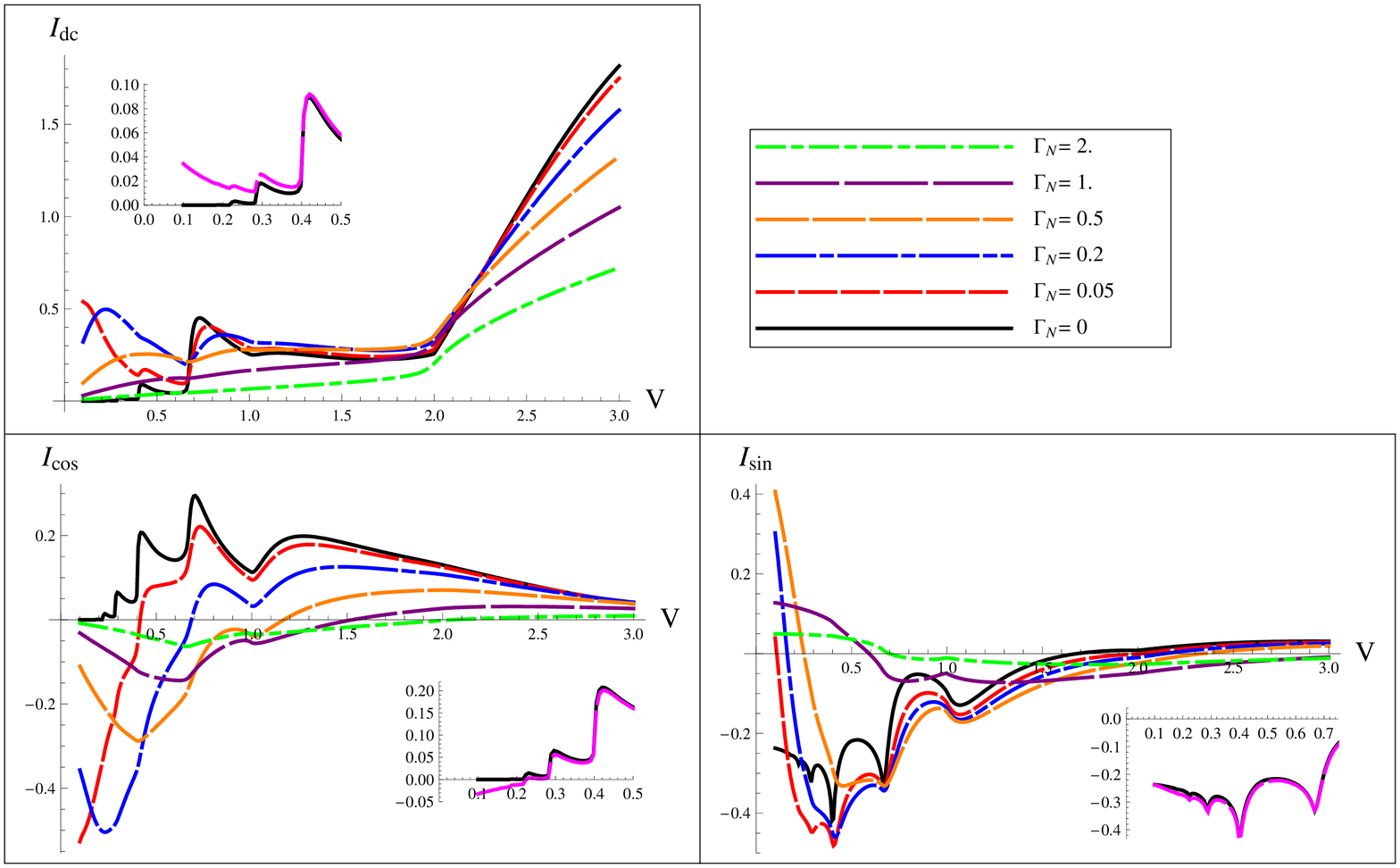}}
\caption{(Color online) Effect of a normal lead. DC (top panel), cosine first harmonic (lower left panel), and 
sine first harmonic (lower right panel) of the MAR current 
for a resonant level at $\Gamma_S=0.2$, with different couplings to the 
normal lead $\Gamma_N$, and with $\epsilon_0=0$. Insets: comparison of the DC 
current and corresponding harmonics for $\Gamma_N=0$ (black) 
and $\Gamma_N=0.02$ (red). 
\label{figure4}}
\end{figure*}

In the top panel of Fig. \ref{figure4} the DC current is plotted, 
for increasing values of the coupling $\Gamma_N$ to the normal 
lead. One notices that a moderate amount ($\Gamma_N=0.05$) of coupling is 
sufficient to provoke an important deviation with respect to the 
case in the absence of coupling: at $\Gamma_N=0.05$ the DC current displays 
a substantial peak close to $eV=.1\Delta$, and we are not able to show the decay 
of this current for $V\to 0$ because of numerical limitations. 
In the insets of the panels of Fig. \ref{figure4}, we compare
the case $\Gamma_N=0$ and $\Gamma_N=0.02$ to show that the 
strong departure from the situation with no coupling is indeed gradual.  
Further increasing the coupling to $\Gamma_N=0.2$ reduces the peak 
amplitude, and shifts both this peak and the overall MAR structure slightly to the 
right.
Eventually, at $\Gamma_N=0.5, 1., 2.$, the low voltage peak and the MAR onset structure at
$eV=2\Delta/3$ are lost, the amplitude of the DC current is reduced, but 
the quasiparticle onset is clearly identifiable.
By extrapolation of our data at $V\to 0$, we also observe (for $\Gamma_N=0.2-2.$) 
that to a very good accuracy the decay of the DC current at low voltage 
is linear. However it is known that in the coherent regime, this decay at low voltages 
is exponential. This constitutes a signature that the gradual coupling of the dot
to the normal lead allows to describe a crossover from coherent 
to incoherent MAR. The enhancement of the low voltage current
for small $\Gamma_N \leq \Gamma_S$ remains a puzzle.    

A similar anomaly occurs at low voltages for the cosine harmonic when 
$\Gamma_N$ is switched on. From the left panel of Fig. \ref{figure4}
and its inset, we see a sharp negative value dip at small voltage and small $\Gamma_N$,
 which becomes more shallow and moves towards larger voltages when the coupling to the normal probe increases.
 For $\Gamma_N=2$, the amplitude 
of the cosine harmonic is strongly suppressed compared to its value 
in the absence of normal lead coupling. For $\Gamma_N=0.2-2.$, 
the extrapolation to low voltage of our results show 
a linear behavior in voltage. The presence of the normal lead 
seems to restore some of the features found for the QPC 
regime (negative sign) because the resonance is broadened. However, 
this broadening also erases most of the information on the MAR onsets
because phase coherence is gradually destroyed. 
Also, note that some dip is formed at $eV=\Delta$, although no 
noticeable structure was found in the DC current.

For the sine harmonic, we saw in Sec. \ref{resonant_section}
for the resonant case (with $\Gamma_N=0$) that the current seemed to saturate
at negative values for $V\to 0$. 
In the presence of coupling to the normal lead however (Fig. \ref{figure4}, lower right panel), 
this is no longer true: upon reduction of the voltage, this harmonic 
switches from negative to positive. The coupling to the normal lead thus restore the equilibrium
population of the Andreev level for small $V$.
In Sect. \ref{josephson_section}, we will show that 
the extrapolation of these curves at $V=0$ truly corresponds to 
the Josephson current. 
In contrast to the cosine harmonic, significant MAR 
structures are observed up to $\Gamma_N=0.5$, but large $\Gamma_N$ eventually 
sets the sine harmonic to zero.      

In summary, the normal lead acts like a dephasing probe, or, equivalently
it is subject to the reverse proximity effect which destroys superconducting 
correlations on the dot. 
First, the coupling to the normal lead tends 
of course to smoothen the current features,
but also it brings a reduction of the supercurrent harmonics because 
of a lack of phase coherence. The signatures of MAR are due to (many) 
round trips of electrons and holes between the superconducting leads, but 
if such carriers are brought in contact with a reservoir, only the DC current
survives, while the AC current vanishes. For large coupling 
to the normal reservoir,  
the DC current resembles that of two 
normal metal-superconductor junctions in series, and the linear 
voltage dependence of the DC current is an illustration of this fact.
Also, note that the current reduction in the presence of a dephasing lead for 
a single normal metal-superconductor junction was discussed in 
Ref. \onlinecite{mortensen}. Any resonant feature
associated with such a junction gets broadened by the presence of
the dephasing probe, and this leads to an effective reduction of
the conductance but also of the Fano factor. 
Our microscopic approach of the DC supercurrent and its first 
harmonics allows to describe the full crossover from coherent 
transport to incoherent transport.  

\section{Josephson current}
\label{josephson_section}

In this section, we compute to Josephson current for the setup including the normal lead.
This allow us to check that the non equilibrium current extrapolated at $V=0$ corresponds to 
the Josephson current.This check was so far performed for a point contact only.\cite{averin} 
 Here we extend it to the case of a resonant level, in
the presence of the normal lead. Note that this geometry (Josephson junction with 
a normal lead connected to the central region) was studied previously
\cite{bagwell} in the context of scattering theory, to probe 
how the Josephson current is affected by current injection from the normal lead. 
Here we assume no current injection, the normal lead plays a passive role. 
    
In order to compute the Josephson current, 
we use the  imaginary-time (Matsubara) path-integral approach to calculate the 
partition function $Z$ and then the Josephson current.
The Hamiltonian is the same as in Sec. \ref{model_hamiltonian},
except that a constant phase difference is imposed between the superconductors.
The lead degrees of freedom are integrated out in the same manner as before.   
This yields the partition function
($\omega$ is a Matsubara fermionic frequency):
\begin{equation}
Z = \int {\cal D} \bar{d} \, {\cal D} d \,  e^{-S} ~,~~~
S = \beta^{-1} \sum_\omega \, \bar{d}_\omega  {\cal L}_{\omega}  d_{\omega} ~,
\end{equation}
with an effective (Euclidean) Lagrangian [matrix in Nambu space] of the dot
\begin{equation}
{\cal L}_\omega = - i \omega \left( 1 + \alpha_\omega \right) + \epsilon_0 \sigma_z +
\alpha_\omega \Delta \cos \frac{\phi}{2} \, \sigma_x - i \, {\rm sign} \, \omega \, \Gamma_N
~,~~~ \alpha_\omega = \frac{2 \Gamma_S}{\sqrt{\omega^2 + \Delta^2}} ~,
\end{equation}
where $\epsilon_0$ is the dot level measured from the chemical potential of
the superconducting leads, $\Gamma_S$ and $\Gamma_N$ are the tunnel widths of the dot due to 
its coupling to
the left/right superconducting lead and the normal electrode, respectively,
$\phi$ is the superconducting phase difference.

With these notations, the partition function reduces to a product of $2\times 2$ determinants:
\begin{equation}
Z=\prod_{\omega_n} \det {\cal L}_{\omega_n}
\end{equation}
with $\omega_n=(2n+1)\pi/\beta$, 
and the Josephson current is obtained from its logarithmic derivative:
\begin{equation}
J(\phi) = -\frac{2}{\beta}\frac{\partial}{\partial \phi}\ln Z\nonumber = 
-\frac{\Delta^2}{\beta} 
\sum_{\omega_n}\frac{\alpha_n^2}{\det {\cal L}_{\omega_n}}\, \sin\phi ~.
\end{equation}

Note that this expression is similar to that found in Ref. \onlinecite{benjamin}, 
except that the escape rate $\Gamma_N$ now appears in the determinant. 
The sum is computed numerically and the sine harmonic of $J(\phi)$ is extracted.
At the same time, we extrapolate the sine harmonic of the non-equilibrium current 
(Sec. \ref{results_for non_interacting dot_section} C) to zero voltage, and we compare the two.  

\begin{figure}
\centerline{\includegraphics[width=8cm]{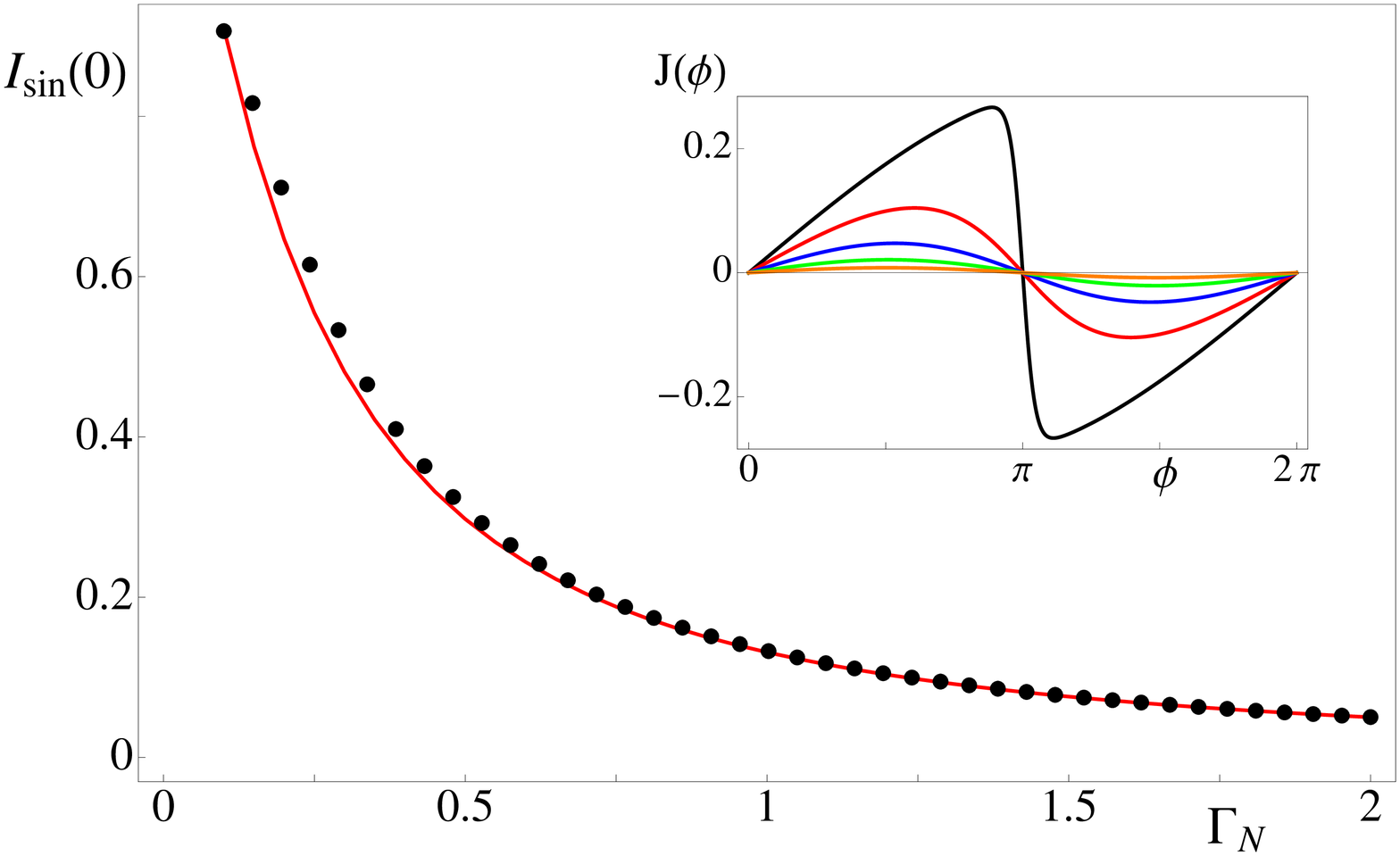}}
\caption{(Color online) Comparison between the first sine harmonic of the Josephson current $J(\phi)$ (full curve)
and the extrapolation at zero voltage of the first sine harmonic of the MAR current (circles).   
Inset: current phase relationship of the Josephson current
for $\Gamma_N = 0, 0.2, 0.5, 1.0, 2.0$ (from larger to smaller amplitude). For both plots the
parameters  are $\Gamma_S=0.2$ and $\epsilon_0=0$
\label{figure5}}
\end{figure}

Fig. \ref{figure5} shows the comparison, for $\Gamma_S=0.2$ (as in section~\ref{Mar_gammaN_section}). The
extrapolation of the out-of-equilibrium case has been obtained by computing the sine harmonics at $V=0.1$ and $V=0.05$.
We see that the agreement between the sine harmonic of the Josephson current and 
the extrapolated MAR harmonics is very good (there is no adjustable parameter), for values of $\Gamma_N$ between 0.1 and 2.0.
Note that the precision of the extrapolation increases when $\Gamma_N$ increases (as $I_{sin}$ becomes flatter near $V=0$ when
$\Gamma_N$ is larger, see Fig.~\ref{figure4}), which explains that the agreement is not totally perfect for the smaller values
of $\Gamma_N$.
Going to values of $\Gamma_N$ smaller than 0.1(not shown) leads to some larger discrepancy between the two types of calculations 
because of the presence of resonant transmission at $\epsilon_0$ with unit transmission.   
The inset of Fig. \ref{figure5} shows the current phase relationship of the Josephson current. 
For large $\Gamma_N$ the current phase relationship becomes essentially sinusoidal, as 
for a tunnel junction. The presence of the normal lead randomizes the phase of 
electrons and holes, and it is equivalent to reducing the 
transmission probability: we have computed the current phase relationship 
at $\Gamma _N=0$, while displacing the dot level $\epsilon_0\in [0.,4.]$ to reduce the transparency 
of the junction (not shown), and found quantitative agreement between this situation 
and that where the normal lead is present.

\section{Conclusion}

We have considered non-equilibrium transport through a quantum dot
sandwiched between superconducting leads in the subgap regime.
The DC and the first harmonics of the current at the Josephson 
frequency were computed, both in the presence and in the absence
of a normal lead connected to the dot. 
 
When the dot level is shifted from the middle of the bias voltage window, 
and the tunneling rates to the superconducting leads are large 
compared to the gap, the current resembles closely that of a superconducting 
QPC. This is explicit for the DC current, but some difference remain for the 
harmonics. Below the quasiparticle onset, the cosine harmonic clearly identifies 
with the QPC case, but it deviates from the latter at higher voltages. 
The sine harmonic shows qualitative agreement with the QPC situation, except that 
the overall signal is shifted downwards. This, for instance, is responsible for a 
sign change in  the sine harmonic at large voltage, while such change is not 
observed a for a QPC. This is attributed to the fact that 
the resonance features of the junction cannot be avoided for the parameter chosen, 
and they affect more the current harmonics than the DC current. 
  
We have also reproduced the DC current of a resonant level in the absence 
of interactions, and the harmonics of the current 
have been obtained. For this situation, we recover the known
fact that only the odd MAR processes survice the DC current. These onsets are also present 
for the harmonics, but the more dramatic effect is that the sign of the 
current harmonics is reversed with respect to the QPC case.

In the presence a coupling to a normal lead, the Keldysh formalism can be
extended by the addition of a self energy in the dot Green's function, with
appropriate Fermi/tunneling phase factors requiring that no current flows in the
dot. So far most of the works on MAR have nevertheless focused on
either on the coherent or the incoherent MAR, and little has been
said about the gradual change from one regime to the other (see although Ref. \onlinecite{Samuelsson2002}). This
has been one of the main focus of this study. The gradual coupling to the normal 
lead are at first not dramatic for the DC current.
The coupling to the lead first smoothes out the MAR features, but as it is increased
a linear (rather than exponential) dependence of the current is found at low voltages, 
and only the quasiparticle onset survives. This DC current voltage 
characteristics corresponds to two normal metal-superconductor junctions in series
with a total lack of phase coherence between the two. 
The harmonics do display a rather dramatic behavior: upon increasing the coupling 
to the normal lead from zero, the amplitude of both harmonics rises sharply, and it saturates 
when the tunneling rate to the normal and the superconducting leads are comparable. 
The harmonics retain some structure at the predicted onsets for larger
coupling than the DC harmonics.

An important check of this work was to compare the first harmonic of the Josephson 
current with the extrapolated value of the MAR sine harmonic at zero voltage. 
We found that in the presence of the normal lead, the agreement is excellent, 
but in the absence of so called dephasing ($\Gamma_N=0$), the resonant features in the 
MAR process render the comparison difficult.

This work could be extended in several directions, some in a 
straightforward manner: The Keldysh formulation 
which was adopted here is quite flexible and can be adapted to 
treat multi-terminal devices, or alternatively multilevel dots 
(which are likely to give rise to several resonances in the subgap region) or 
more complicated structures described by tight binding Hamiltonians.    

Other perspectives of this study concern the inclusion of interactions.
Few work include so far electronic correlations on the dot in the 
calculation of the DC current\cite{avishai,dellanna}. 
The approach chosen in Ref. \onlinecite{avishai} requires a minimal 
Coulomb interaction parameter in order to trigger a departure from the non interacting regime.
There, interactions are shown to reduce drastically the DC-current, to shift 
and to damp out the structure found at the MAR onsets.
The approach of Ref. \cite{dellanna} is complementary as it treats weak interactions
in a systematic perturbation theory approach: there, interactions are shown to lead 
to a small enhancement of the DC current. 

At the same time, electron-phonon interactions have been included
perturbatively \cite{zazunov_egger} in the calculation of the DC MAR current.
While inelastic scattering can provide a source of dephasing, the
choice of the low temperature regime and the weak electron-phonon coupling
assumption resulted there in a current voltage characteristics 
which is only weakly affected by the vibrations: no dissipation 
associated with these vibrations was noticed. At higher
temperatures and stronger phonon coupling, phonon scattering could
provide a definite source of decoherence.

In some sense, the simple model presented here for the inclusion of 
decoherence could serve as a point of comparison 
for a more complete study 
of a dot strongly coupled to phonons out of equilibrium. 

\vspace{1.cm}
\appendix  

\section{Numerical implementation}
\label{sec:appendix}
The numerical implementation follows directly from equations~(\ref{eq01}) to (\ref{eq:dressedG}). 
The matrix inversion giving the dressed Green function (Eq.~(\ref{eq:dressedG})) can be done
once the sum $\sum_{n,m}$ over the frequency domains has been truncated.
We have chosen a constant width $E_c$ for each domain, defining a $V$-dependent $n_{max} = E_c / e V$.
It is clear that the smallest values of $V$ are the most expensive to obtain numerically, which explains
why our results are limited to $eV/\Delta>0.1$. We have taken great care to check
that our results do not suffer from convergence problem due to the truncation.
In practice, we have chosen $E_c =16 \Delta$, apart from some special cases in the QPC limit
where larger $E_c$ were needed (see the discussion in the QPC section).  Note that for most 
of the curves shown in the figures, convergence could be obtained with a much lower value of $E_c$.
Once  the dressed Green function $G_{nm}$ is computed, the current is obtained by the numerical
integration over the fundamental domain $[-V/2,V/2]$ given in Eq.~(\ref{eq01}). 

We have used the following Green function sum rules\cite{kamenev,zazunov_egger} as independent checks of our method:
\begin{eqnarray}
{\rm Tr} ~\{\tau_y \sigma_z G(t,t) \} & = & 0 ~, \\
{\rm Tr} ~\{ -\tau_y G(t,t) \} & = & 2 ~,
\end{eqnarray}
where $\sigma_z$ acts in Nambu space, $\tau_y$ acts in Keldysh space, and the trace 
is taken on both Nambu and Keldysh space.

We found that the first sum rule was always satisfied to large precision within our truncation scheme. The second sum rule
is only approximatively satisfied, and gives useful information on the quality of the truncation. Typically,
it has a value larger than 1.9 for most of the results shown, but can be somewhat lower when $\Gamma_S$ increases.
However, we have found that the computed current does not depend critically on the precise
validation of this sum rule (e.g. in the very large $\Gamma_S$ limit, correct value of the current can be obtained even with this sum 
rule giving values as low as 1.6).

\acknowledgments

T.J. and T.M acknowledge financial support from ANR PNANO grant ``Molspintronics''.
K.B. benefited from the Ecole Normale-Landau Institute exchange program.
V.S. thanks the Centre de Physique Th\'eorique (UMR 6207 of CNRS) for its
hospitality.



\begin{thebibliography}{99}

\bibitem{KBT}
T.M. Klapwijk, G.E. Blonder, and M. Tinkham, Physica (Amsterdam)
109B-110B, 1657 (1982).

\bibitem{arnold}
G.B. Arnold, J. Low Temp. Phys. {\bf 68}, 1 (1987).

\bibitem{burstein} B. N. Taylor and E. Burstein, Phys. Rev. Lett. {\bf 10}, 14 (1963); 
 J.R. Schrieffer and J.W. Wilkins, Phys. Rev. Lett. {\bf 10}, 17 (1963).

\bibitem{bratus} E.N. Bratus, V.S. Shumeiko, and G. Wendin,
Phys. Rev. Lett. {\bf 74}, 2110 (1995).

\bibitem{averin}
D. Averin and A. Bardas, Phys. Rev. Lett. {\bf 75}, 1831 (1995).

\bibitem{cuevas}
J.C. Cuevas, A. Martin-Rodero, and A.L. Yeyati, Phys. Rev. B {\bf
54}, 7366 (1996).


\bibitem{mar_point_contacts_exp_current}
N. van der Post, E.T. Peters, I.K. Yanson and J.M. van Ruitenbeek,Phys. Rev. Lett., {\bf 73}, 2611 (1994);
E. Scheer, P. Joyez, D. Esteve, C. Urbina,* and M. H. Devoret,
Phys. Rev. Lett. {\bf 78}, 3535 (1997);
E. Scheer, N. Agra\"\i{i}t, Juan Carlos Cuevas,
A. Levy-Yeyati, B. Ludophk, A. Martin-Rodero,
G. Rubio Bollinger, J. M. van Ruitenbeek,
C. Urbina, Nature, {\bf 394}, 154 (1998).

\bibitem{mar_point_contacts_exp_noise}
R. Cron, M. F. Goffman, D. Esteve, and C. Urbina
Phys. Rev. Lett. {\bf 86}, 4104 (2001).

\bibitem{diffusive MAR_experiments}
J. Kutchinsky, R. Taboryski, T. Clausen, C. B. Sørensen, A. Kristensen,
 P. E. Lindelof, J. Bindslev Hansen, C. Schelde Jacobsen, and J. L. Skov, 
Phys. Rev. Lett. {\bf 78}, 931 (1997);
T. Hoss, C. Strunk, T. Nussbaumer, R. Huber, U. Staufer, and C. Sch\"onenberger,
Phys. Rev. B {\bf 62}, 4079 (2000);
C. Hoffmann, F. Lefloch, M. Sanquer, and B. Pannetier Phys. Rev. B
{\bf 70}, 180503(R) (2004).

\bibitem{diffusive_MAR_theory}
E. V. Bezuglyi, E. N. Bratus, V. S. Shumeiko, G. Wendin, and H. Takayanagi,
Phys. Rev. B {\bf 62}, 14439 (2000);
J. C. Cuevas, J. Hammer, J. Kopu, J. K. Viljas, and M. Eschrig,
Phys. Rev. B {\bf 73}, 184505 (2006);
G. Niebler, G. Cuniberti, and T. Novotny, Supercond. Sci. Technol. {\bf 22}, 085016 (2009).

\bibitem{alfredo} A.L. Yeyati, J.C. Cuevas, A. L{\'o}pez-D{\'a}valos,
and A. Martin-Rodero, Phys. Rev. B {\bf 55}, R6137 (1997).

\bibitem{johansson}
G. Johansson, E.N. Bratus, V.S. Shumeiko, and G. Wendin, Phys.
Rev. B {\bf 60}, 1382 (1999).

\bibitem{schonenberger}
M. R. Buitelaar, T. Nussbaumer, and C. Sch\"onenberger, Phys. Rev. Lett. {\bf 89}, 256801 (2002);
M. R. Buitelaar, W. Belzig, T. Nussbaumer, B. Babic, C. Bruder, and C. Sch\"onenberger,
Phys. Rev. Lett. {\bf 91}, 057005 (2003).

\bibitem{experiments_MAR_ quantum_dots}
H. I. Jorgensen,1, K. Grove-Rasmussen, T. NovotnyŽ, K. Flensberg, and P. E. Lindelof
Phys. Rev. Lett. {\bf 96}, 207003 (2006);
T. Sand-Jespersen, J. Paaske, B. M. Andersen, K. Grove-Rasmussen, H. I. Jørgensen, M. Aagesen, C. B. Sørensen,
 P. E. Lindelof, K. Flensberg, and J. Nygård, Phys. Rev. Lett. {\bf 99}, 126603 (2007).

\bibitem{Samuelsson2002}  
P. Samuelsson, G. Johansson, \AA. Ingerman, V. Shumeiko, and  G.Wendin, Phys. Rev.  B {\bf 65}, 180514(R) (2002).

\bibitem{buttiker_dephasing}
M. B\"uttiker, Phys. Rev. B {\bf 33}, 3020 (1986).

\bibitem{beenakker_buttiker} C. W. J. Beenakker, and M. B\"uttiker,
Phys. Rev. B {\bf 46}, 1889 (1992).

\bibitem{konig}
M.G. Pala, M. Governale, and J. K\"onig, New J. Phys. {\bf 9}, 278 (2007).

\bibitem{zazunov_feinberg} A. Zazunov, D. Feinberg, and T. Martin,
Phys. Rev. Lett. {\bf 97}, 196801 (2006).

\bibitem{zazunov_egger}
A. Zazunov, R. Egger, C. Mora, and T. Martin, Phys. Rev. B {\bf
73}, 214501 (2006).

\bibitem{mw}
Y. Meir and N.S. Wingreen, Phys. Rev. Lett. {\bf 68}, 2512 (1992).

\bibitem{bratus1997}
E.N. Bratus', V.S. Shumeiko, E.V. Bezuglyi, and G. Wendin, Phys. Rev. B {\bf 55}, 12666 (1997).

\bibitem{glazman_matveev}
  L. I. Glazman and K. A. Matveev, Pis'ma Zh. Teor. Fiz. \textbf{49}, 570 (1989) [JETP Lett. \textbf{49}, 659 (1989)].

\bibitem{spivak_kivelson}
  B. I. Spivak and S. A. Kivelson, Phys. Rev. B \textbf{43}, 3740 (1991).

\bibitem{benjamin}  C. Benjamin, T. Jonckheere, A. Zazunov, and T. Martin, Eur. Phys. J. B 57, 279-289 (2007). 

\bibitem{michelsen2008} J. Michelsen, V.S. Shumeiko, and  G. Wendin, Phys. Rev. B {\bf 77}, 184506 (2008).

\bibitem{beenakker_van_houten}
 C.W.J. Beenakker, and H. van Houten,
in {\it Single-Electron Tunneling and Mesoscopic Devices}, H. Koch and H. Lubbig eds. (Springer, Berlin, 1992). 

\bibitem{wendin1996} 
G. Wendin and V.S. Shumeiko, Superlattices and Microstructures {\bf 20}, 569 (1996).

\bibitem{mortensen} 
N.A. Mortensen, A.-P. Jauho, and K. Flensberg,
Superlattice Microstr. {\bf 28}, 67 (2000).

\bibitem{bagwell} Li-Fu Chang and Philip F. Bagwell, Phys. Rev. B {\bf 55}, 12678 (1997);
P. Samuelsson, V.S. Shumeiko, and G. Wendin, Phys. Rev. B,  {\bf 56}, R5763 (1997);P. Samuelsson, J. Lantz, V.S. Shumeiko, and G. Wendin, Phys. Rev. B {\bf 62}, 1319 (2000).

\bibitem{avishai}
Y. Avishai, A. Golub, and D. Zaikin, Phys. Rev. B {\bf 63}, 134515
(2001).

\bibitem{dellanna}
L. Dell'Anna, A. Zazunov, and R. Egger
Phys. Rev. B 77, 104525 (2008).

\bibitem{kamenev}
A. Kamenev, in {\sl Nanophysics: Coherence and Transport} Les
Houches session LXXXI, ed.~by H. Bouchiat, S. Guéron, Y . Gefen, G. Montambaux and J. Dalibard (Elsevier,
2005).




\end{thebibliography}
\end{document}